\documentstyle[12pt]{article}
\begin{document}
\title{Chaos as a Source of Complexity and Diversity in Evolution}
\author{
        Kunihiko KANEKO \\
        {\small \sl Department of Pure and Applied Sciences}\\
        {\small \sl University of Tokyo, Komaba, Meguro-ku, Tokyo 153,
JAPAN}
\\}
\date{}
\maketitle
\begin{abstract}
The relevance of chaos to evolution is discussed in the context of the
origin and maintenance of diversity and complexity.  Evolution to
the edge of chaos is demonstrated in an imitation game.
As an origin of diversity, dynamic clustering of identical
chaotic elements, globally coupled each to other, is briefly reviewed.
The clustering is extended to nonlinear dynamics on hypercubic lattices,
which enables us to construct a self-organizing genetic algorithm.
A mechanism of maintenance of diversity, ``homeochaos", is given in an
ecological system with interaction among many species.  Homeochaos
provides a dynamic stability sustained by high-dimensional weak chaos.
A novel mechanism of cell differentiation is presented, based on
dynamic clustering.  Here, a new concept --  ``open chaos" --  is
proposed for the instability in a dynamical system with growing degrees of
freedom.  It is suggested that studies based on interacting chaotic elements
can replace both top-down and bottom-up approaches.

\end{abstract}
\section{Complexity, Diversity, and Emergence}
Why are we interested in the effort to create "life-like" behavior in
computers? The answers can be diverse, but my interest in
such artificial biology lies in the construction of systems exhibiting
the emergence and maintenance of complexity and diversity, in order
to understand the evolution of the complex "society" of life.
This problem is not so trivial, indeed.  It is often difficult
to conclude that a system's emergent complexity is somewhat beyond
that which would be expected on the basis of the rules explicitly
implemented within a model \cite{GEB}. Often, what people call
``emergent" behavior comes from the lack of a full understanding of what
is implied by the rules implemented in the model.
In evolution, there is a stage of the emergence of novel features
as well as a stage of slow-scale change of existing features.
Gradual evolution after the emergence of
a novel feature is often studied analytically with the use of stochastic
differential equations, as, for example, is demonstrated by
the neutral theory of evolution \cite{Kimura}.
The ``origin" of features, on the other hand, is often a difficult
problem to solve analytically.  The origins of life, eukaryotes, multi-cellular
organism, germ-line segregation, and sex are examples of the emergence of
such novel features. For such problems, we require a mechanism for how
complex, higher-level behavior emerges from low-level interactions,
without the implementation of explicit rules for such emergence.
Such emergence, we believe, occurs through strong nonlinear interactions
among the agents at the lower level.
Nonlinear interaction among agents often leads to chaotic behavior, which,
we believe, can cause ``aufheben" (German terminology of dialectic
philosophy) to higher level dynamics, by which lower level conflicts are
resolved.
In this overview, we try to demonstrate that chaos is relevant to the
emergence and maintenance of complexity and diversity.
Chaos is the most universal mechanism to create
complexity from simple rules and initial conditions.
As will be seen, chaos can be a source of diversity: identical elements
differentiate through chaotic dynamics. Through a dynamical process
with instability, chaos also has the potentiality to create a higher-level
dynamics.
Problems we address in the present overview are as follows;
(i) evolution to complexity, (ii) sources of diversity, (iii) maintenance of
diversity, and (iv)  successive creation of novelty and open-ended
evolution to diversity.
In \S 2, evolution to the edge of chaos, a complex state between
chaos and order (a window), is studied with the use of an imitation game.
An explicit example of the evolution of complexity is given.
General concepts in globally coupled dynamical systems are briefly given
in \S 3, including the dynamic clustering of synchronization,  hidden
coherence, and chaotic itinerancy.
In \S 4 these novel concepts are applied to dynamical systems
on a hypercubic lattice, which enables us to construct spontaneous genetic
algorithms.  A new concept -- ``homeochaos" -- representing
dynamical stability involving weak chaos with many degrees of freedom,
is given in \S 5, as well as its relevance to various biological networks.
Homeochaos (to be contrasted with "homeostasis") can provide for the
maintenance of diversity. In \S 6, the concept of clustering in
globally coupled maps is extended to the problem of
cell division and differentiation.
Here the novel concept of ``open chaos" is demonstrated in a system with
growing degrees of freedom.  Open chaos leads to the formation of
disparities in activities among cells, leading to the emergence of diversity
and novelty. In \S 7 we show the advantage of our approach over top-
down and bottom-up approaches.
\section{Edge of Chaos in an Imitation game: Chaos as a source of
Complexity}
The increase of complexity through evolution is believed to be seen in
many biological systems, not only in the hierarchical
organization in genotypes and phenotypes but also in animal behavior and
communication.
A direction for the increase of complexity has recently been discussed as
"evolution to the edge of chaos" \cite{Packard}, since complexity is
believed to be large at the border between order and chaos
\cite{Grass,Crutch,Chris,Kauf2,KK-PD}. However, there has been no clear
simple example providing evidence of evolution to the edge of chaos, in
the exact sense of dynamical systems theory.  Chaos
is defined only on dynamical systems with a continuos state,
and is not defined for discrete-state systems such as cellular automata,
which have been adopted for studies of the edge of chaos so far.
Recently Suzuki and the author have presented an example of evolution to
the edge of chaos by introducing a simple model for an imitation game of a
bird song \cite{KKJS}.
A bird song, for example, is known to increase its complexity through
evolution and development (with more repertoire made up from
combinations of simple phrases)\cite{Catch}. A bird with a complex song
is stronger in defending its territory \cite{Catch,Krebs}.
Based on this observation of a function of bird song for the
defense of territory, we have introduced an imitation game \cite{KKJS},
in which the player who imitates the other's song better wins the game.
As a "song", a time series generated by a simple
mapping $x_{n+1}=f(x_n)=1-ax_n ^2$ (the logistic map) is adopted.
As is well known the attractor of
the map shows a bifurcation sequence from a
fixed point, to cycles with periods 2,4,8,..., and
to chaos as the parameter $a$ is increased \cite{logistic}.
Here the parameter value $a$ is assumed to be
different for each individual "bird".  By this choice of a song generator,
one can examine whether a song evolves towards the edge of chaos.
Each bird player $i$ chooses an initial condition, so that the time series of
its own dynamics $x_{n+1}(i)=f_i (x)=1-a(i) x_n (i)^2$ can
imitate the song of another player.  For ``preparation" of an initial
condition, the player $i$ uses a feedback from the other player $j$'s song
by $x_{n+1}(i)=f_i [(1-\epsilon)x_n (i)+\epsilon x_n (j)]$ over a number of
given time
steps $t_{trs}$.  Of course, birds have to choose initial conditions for
starting the above feedback process, and also
for singing, and the result of a game can depend on this choice of initial
conditions \cite{KKJS}.  Here we assume that they choose initial conditions
randomly over $[-1,1]$.  Thus the game is probabilistic, although
``strong" players (to be discussed) often win against ``weak ones" with
probability close (or equal) to one \cite{KKJS}.
Repeating the imitation process, the
distance between two songs $D(j,i)=\sum_{n=t_{trs}}^{t_{trs}+T} (x(j)-
x(i))^2$  is measured. By changing the role of the player $i$ and $j$,
$D(j,i)$ is measured.
If $D(i,j)<D(j,i)$, the player $i$ imitates better than $j$, thus being
the winner of the game, and vice versa.
By reproducing the players
according to their scores in the game, and by including mutation of the
parameter $a$ \cite{GA}, we have examined the dynamical states
to which the songs evolve.
Temporal evolution of the average of the parameter $a$ over all players
shows successive plateaus, until it reaches $a \approx 1.94$, where it then
remains. Plateaus corresponding to period-doubling bifurcation points or to
the edge between periodic windows and chaos are observed successively.
In Fig.1, the average score of players is plotted as
a function of $a$.  The score has a peak at the edge between
chaos and windows for stable periodic cycles.
To study evolution to the edge of chaos,
we have also plotted the score of birds as a function of the Lyapunov
exponent $\lambda $ for the dynamics $x \rightarrow f(x)$ (see Fig.2).
Indeed, the score has a broad peak around $\lambda = 0$,
corresponding to the edge of chaos.
Thus, evolution of a song towards the edge of chaos is observed.
The final value of $a \approx 1.94$ corresponds to
the borderline between a periodic window (of period 4)  and chaos.
Besides evolution to the edge of chaos, it should be noted that
the "edge" reached by evolution lies between a periodic window and chaos.
At a window, the dynamics show chaotic transients before attraction to a
stable cycle.  Here, a variety of unstable cycles coexist
\cite{logistic}, which provide for a larger variety of dynamics, as
transients. Transient chaos is important
for the adaptation to a wide range of external dynamics.
Evolution {\bf to the edge of windows} may be a more robust and
important concept than evolution to the edge of chaos.
Escape from imitation could be a trigger for the
evolution of complexity in many fields, such as in the evolution of
Batesian mimicry \cite{Maynard}, where
one of the groups can survive better by imitating the
pattern of another group, while the second group's advantage in
survival is lost if it is not distinguished well from the first.
The increase of complexity of the patterns of some butterflies
may be due to this ``imitation" pressure.
Another possible application is seen
in the evolution of a communication code only
within a given group.
Studies of the evolution of such signals will be important in the future,
from the viewpoint of complexity via chaotic dynamics.
\section{Key Concept for the Origin of Complexity and Diversity:
Dynamic Clustering in Networks of Chaotic Elements}
To study the emergence of diversity, we need a mechanism
by which identical elements differentiate into different groups
spontaneously.  Networks of chaotic elements, globally coupled to
each other, provide an example of such a mechanism.

In many biological networks, the interaction among elements is
not local but global. The simpliest case of global interaction is studied
as a ``globally coupled map" (GCM) of chaotic elements.
An example is given by
\begin{equation}
x_{n+1}(i)=(1-\epsilon )f(x_{n}(i))+\frac{\epsilon }{N}\sum_{j=1}^N
f(x_{n}(j))
\end{equation}
where $n$ is a discrete time step and $i$ is the index of an
element ($i= 1,2, \cdots ,N$ = system size), and $f(x)=1-ax^{2}$ \cite{GCM}.
Without coupling  (i.e., for $\epsilon =0$), each element
shows chaotic behavior if $a$ is large enough.
The model is a mean-field-theory-type extension of coupled map
lattices (CML).  The above dynamics consists of parallel nonlinear
transformation with a feedback from the ``mean-field"  \cite{equiv}.
In real biology, elements are not necessarily identical.
The reason we start from identical elements is that we are interested in
the origin of differentiation and diversity. That is, we are interested in the
question: How can a set of identical units evolve to groups with different
(dynamical) states?
Through interaction, some elements oscillate synchronously,
while chaotic instability gives a tendency for the destruction of coherence.
Attractors in GCM are classified by the number of synchronized
clusters $k$ and the number of elements within each cluster $N_k $.
Here, a cluster is defined as the set of elements in which $x(i)=x(j)$.
Identical elements split into clusters with
different frequencies, phases, or amplitudes of oscillation.
Each attractor is coded by the
clustering condition $[k,(N_1 ,N_2 ,\cdots,N_k )]$.
In a globally coupled chaotic system in general,
the following phases appear successively with the increase of
nonlinearity in the system \cite{GCM}:
(i) {\bf Coherent phase}: A coherent attractor ($k=1$)
has occupied (almost) all basin volumes.
(ii) {\bf Ordered phase}: Attractors ($k=o(N)$) with few clusters
have occupied (almost) all basin volumes.
(iii) {\bf Partially ordered phase}: Coexistence of attractors
with many clusters ($k=O(N)$) and attractors with few clusters.
(iv) {\bf Turbulent phase}:  All attractors have many clusters ($k=O(N)$;
in most cases $k \approx N$).
In the turbulent phase,
although $x(i)$ takes almost random values almost independently,
there remains some coherence among elements.
The distribution $P(h)$ of the mean field $h_n \equiv (1/N)\sum _j
f(x(j))$,
sampled over long time steps, does not
obey the  law of large numbers \cite{GCM2}.  The emergence of hidden
coherence is a general property in a globally coupled chaotic system.
This hidden coherence may be interesting in relation to
EEG's, where one measures a given average of neuronal (electric) activity.
Although the firing of each neuron is not regular ( i.e., chaotic or random),
the amplitude of some average (EEG) still has a large enough amplitude
of variation to be observed, which may suggest the existence of
hidden coherence as in the above.
In the partially ordered phase, orbits make itinerance over
ordered states via highly chaotic states.  In the ordered
states the motion is partially coherent.
Our system exhibits intermittent change between
self-organization towards a coherent
structure,  and its collapse to a high-dimensional
disordered motion.  This dynamics, called {\sl chaotic itinerancy},
has been found in a model of neural dynamics by Tsuda \cite{Tsuda},
optical turbulence \cite{Ikeda}, and in GCM \cite{GCM}.
Here, a number of ruins of low-dimensional attractors coexist in the
phase space.  The total dynamics consists of residencies at ruins
interspersed with excursions into high-dimensional chaotic states .
In the chaotic itinerancy in GCM, the degree of synchronization between
two elements changes with time. Elements 1 and 2, for example
may be almost synchronized for some time span, until desynchronization
by high-dimensional chaos destroys the relationship.  After some time,
element 1
may be almost synchronized with element 5, for example, and so forth.
Thus, the relationship between elements is dynamically changing.
Indeed, such change of relationship is discussed in neural physiology
\cite{Kruger-Book}.
\section{Clustering in hypercubic coupled maps; self-organizing genetic
algorithms}
Let us discuss an extension of the idea in \S 3 to population dynamics
with mutation. The process of mutation is characterized by a diffusion
process in the space of genes.  If the ``gene" space is represented
by a bit space (such as $i=$0010111, as is often the case for genetic
algorithms \cite{GA}), the
single point mutation process is given by a flip-flop $0 \leftrightarrow 1$
at each position.  Let us represent the population (density)
of each species $i$ by $x(i)$, the mutation process is given by a
diffusion in hypercubic bit space.  When the population dynamics
is represented by $x(i) \rightarrow f (x(i))$, the total dynamics
is given by
\begin{equation}
x_{n+1}(i)=  (1-\epsilon )f (x_n (i))+ \frac{\epsilon}{k} \sum_{j=1}^k
f(x_n(\sigma _j(i))),
\end{equation}
where $\sigma _j (i)$ is a species whose $j$'th bit is different from
the species $i$ (with only one bit difference) ,
and $k$ is the total bit length of species ( total species is $2^k$)
\cite{HCM-com}.
We use a binary representation to denote the lattice here;
for example, site 42 for $k=6$ means the hypercubic lattice point
101010.
The above model is rather close to the model in the last section;
instead of the global coupling in (1), nearest neighbor coupling
on the hypercube is adopted here.
In the model (2), we have again found the formation of synchronized
clusters
as in \S 3 (i.e.,$x_n(i)=x_n(j)$ for two elements $i$ and $j$ in the cluster).
In the present case, the split to clusters is organized according to the
hypercubic structure.  Examples of such clusters follow.
(a) 1 bit clustering
Two clusters with synchronized oscillation are formed.  Each of the clusters
has $N/2=2^{k-1}$ elements, determined by the bit structure.
For example, elements may be grouped into two clusters
with **0*** and **1***, (* means that the symbol there is either one or
zero), each of which has $2^{k-1}$ species. This clustering is formed by
cutting the k-dimensional hypercube by a hyperplane.
In the genetic algorithm \cite{GA}, irrelevant bits are initially determined
as  ``don't care" bits represented by ``\#".  Here, such bits are
spontaneously created with the temporal evolution.
(b) 2 bit clustering
Depending on initial conditions and parameters, both the number of
clusters and the number of
bits relevant to clustering can be larger than in case (a).
An examples is a 2-cluster state with 2 relevant bits by XOR (exclusive-or)
construction. Here the elements split into
the groups (i) **10*** or **01*** and (ii) **00*** or **11***, for example.
(c) parity check clustering
Elements split into two groups according to the parity of the number of
1's in each bt representation.  For example, elements
split into two clusters as follows
(i) 000, 011, 101, 110 and (ii) 001 010, 100, 111, for $k=3$.
The clustering, thus gives a parity check.
It is a hypercubic version of the  zigzag (1-dim) or checkerboard (2-dim)
pattern \cite{KK-PD}.
Besides these examples, attractors with many clusters are
also found.  Most of these states are constructed by
combining the above clustering schemes.
For example, 4 clusters with two relevant bits are
found as a direct product state of case (a).
Here the hypercubic space is cut by two hyperplanes.
Elements split into
four clusters, for example, coded by 01*****, 10*****,
11*****, and 00*****.
More complex examples are reported in \cite{HCM}.
Here we have to note that not all partitions are
possible in the present case.  Even if we start from
an initial condition with an arbitrary clustering,
the synchronization condition ($x(i)=x(j)$
for $i,j$ belonging to the same cluster)
is not satisfied at the next step for
most such initial conditions.
In contrast with the GCM case, not all possible partitions
can be a ( stable or unstable) solution of the evolution
equation.
As is discussed, the present result opens up the possibility
of automatic genetic algorithms.  Relevant bits are
spontaneously formed.  Furthermore we have found a chaotic itinerancy
state, where relevant bits change according to temporal evolution.
In Fig.3, the change of relevant bits for clustering is clearly seen.
At stage A, two clusters are formed by the first bit ( i.e., clusters
$0****$ and $1****$), and the second bit at stage B
($*0***$ and $*1***$), and so on.
With the introduction of  external inputs to each element, it is also possible
to have a clustered state following the external information
\cite{HCM}.  Relevant information is extracted through this process
spontaneously, which is stored as a relevant bit in the clustering.
An application of the present clustering to ``real life" will be found
in the quasispecies of viruses \cite{Eigen}.  As Eigen et al. discuss,
viruses form quasispecies coded in hypercubic space.
By taking account of population dynamics,
the present clustering may give a theoretical
basis for the dynamic and hierarchical grouping of quasispecies.
\section{Maintenance of Diversity and Dynamic Stability : Homeochaos}
As for the evolution to complexity, the notion of the
``edge of chaos" in \S 2 is rather special.  First, it
is provided by a critical state, and should be sustained at
a very narrow region (or at a single critical point) of
the parameter space.  Second, the system is given by
a low-dimensional dynamical system, i.e., with very few degrees of
freedom.
Hence the notion ``edge of chaos" is insufficient to understand
the diversity and complexity of a biological system.
In an ecological system,
many species are under strong nonlinear interaction, and keep
some kind of stability with diversity.  This is not
easily sustained.
We also have to mention that static equilibria with many species
are usually unstable, as studied by May \cite{May} in
a random network model.
Thus it is interesting to search for a dynamical mechanism to
allow for the diversity in a system with interacting population dynamics.
Ikegami and the author have studied a population dynamics model
with interaction among species, mutation, and mutation of
mutation rates \cite{homeo,homeo2}.
In particular, a model with interaction among
hosts and parasites has been studied.
Each species is coded by a bit sequence as in \S 4, whose
fitness has a rugged or flat (neutral) landscape.
The interaction between a host and a parasite
is assumed to depend on the Hamming distance between
their bit sequences.
When the interaction between hosts and parasites is weak,
the mutation rates of species decrease with evolution. The dynamics
of the whole species is reduced
to a direct product of isolated sets of host-parasite population dynamics.
When the interaction is strong, on the other hand,
mutation rates are sustained at a high level,
where many species form a network of population dynamics.
This network consists of species connected by single point
mutations.  Many species are percolated in the gene space.
Note that this network is dynamically sustained.  The population of each
species oscillates chaotically in time.
The oscillation is high-dimensional chaos with small positive
Lyapunov exponents.  ( ``High-dimensional" here means that
the number of positive exponents is large).
If the mutation rate were zero, the dynamics of each species
would be essentially disconnected.  Then some host-parasite pairs
would show strong chaos, while others would show periodic or fixed
point dynamics.  By sustaining a high mutation rate, chaotic instability
is shared by almost all species, leading to weak high-dimensional chaos.
By the term "weak", we mean that the maximum Lyapunov exponent is
close to
zero, and that the
amplitude of oscillation of each species is small.
Our system has a tendency to evolve towards such
weak, high-dimensional chaos.
Here we propose a conjecture that diversity in an evolutionary system
with interaction of many replicating units maintains its dynamical
stability by forming a weak high-dimensional chaotic state,
rather than in a fixed point or in strong chaos.
We have coined the term {\bf homeochaos} for this
homeodynamic state.
The following three points capture the essence of homeochaos:
(i) {\bf Weak Chaos}:
Homeochaos suppresses strong chaos.  The maximal Lyapunov exponent
is positive, but is close to zero.  The oscillation amplitude is
not large.  This weak chaos, for example, is essential to avoid an overly
violent change or extinction in the population dynamics.
(ii) {\bf High-Dimensional Chaos}:
Homeochaos is high-dimensional chaos. There are many positive
Lyapunov exponents, although their magnitude is small, and there are
many degrees of freedom.
(iii) {\bf Dynamic Stability and Robustness against External Perturbations}:
Homeochaos provides dynamic stability for a complex network.
The robustness of homeochaos is easily seen by introducing
an external perturbation to the population dynamics.  If the
population dynamics follows low-dimensional chaos, the amplitude of
population change is sometimes very large.  The oscillation can bring about
a state with very small population size (see Fig.4).
When an external perturbation
is applied at this time, the number of population may go to zero.
On the other hand, the oscillation amplitude is small in homeochaos (see
Fig.4),
since the chaos is very weak.  Thus populations of species fluctuate around
some value far from zero.  Hence species are not easily driven to extinction
by external perturbations.
The above three features are strongly interrelated.
The stability and robustness (iii) are
sustained by the suppression of strong instability
given by (i). By (ii), strong chaotic instability is shared by many modes,
implying the weak chaos per degrees of freedom ( the point (i)).
The point (i) is a feature common with homeochaos and the edge of chaos.
However homeochaos is not sustained at a critical point, but is
more robust against a parameter change.  Also the degrees of freedom are
not discussed in the edge of chaos, but they are
essential to homeochaos.
Remnants of clustering of oscillation are important in sustaining
homeochaos.
Indeed the chaotic itinerancy seen in clustering is sometimes seen in
homeochaos.  The oscillation of some populations of some species form
partial clustering over some time steps.
The connection between homeochaos and clustering
may not be so surprising.  If chaos were too strong, oscillations
of many elements would not keep any relationship, and they would
become
completely desynchronized.  If chaos were completely suppressed,
clustering with few number of clusters would often follow.  To keep weak
and high dimensional chaos, partial clustering with chaotic itinerancy
is the most preferable state.
Homeochaos in the formation of networks will be important
in various levels of biological networks.
Maybe the most straightforward application
will be found in immune networks and quasispecies of viruses.
In the immune system, antibody-antigen interactions are similar to
host-parasite interactions.
An antigen is damaged by "matched" antibodies.  An antibody itself is
damaged by a different class of antibodies, as
Jerne \cite{Jerne} proposed. High mutation rates are sustained for
antibodies, and the concentrations of antibody species oscillate in time.
We note that these features are shared in common with our model above
and its numerical results.
Correspondingly, viruses keep high mutation rates, and the
population dynamics maintains high mutation rates.  The population
dynamics
of quasispecies \cite{Eigen} is of interest from the viewpoint of
homeochaos.
Indeed the term ``quasispecies"  of \cite{Eigen} corresponds to
our term ``meta-species", a symbiotic network connected by mutation in
\cite{homeo}.
At a more macroscopic level, metabolic oscillations of cells may
form homeochaotic dynamics, as will be discussed in the next section.
At a further macroscopic level, physiology and medicine are
the birthplace of the term ``homeostasis".  However, data have recently
been accumulating showing that the healthy state is not one of "stasis", but
rather exhibits an irregular temporal dynamics.  So far, it is hard to
conclude that the dynamics of the healthy state is chaotic.  Possibly this
difficulty may be
due to the high-dimensionality of the dynamics, where no powerful means
of diagnosis from the data is available.  Indeed, low-dimensional
chaos has to date been found to be associated with unhealthy states of the
heart rhythm and EEG.
One possible conjecture is that a healthy state is sustained by
homeochaos rather than homeostasis, since homeochaotic dynamics is
neither too irregular nor too regular.
At the most macroscopic level,
an ecological network can be a candidate for homeochaos.
As Elton has discovered in the forest of England \cite{Elton},
an ecological system
with diversity is robust against external perturbations.
A typical example of a complex ecological network is found among
the species in a tropical rain forest.  The ecology there consists of
a huge number of species, whose population size is often very small.
So far, the dynamics of the population of species in a complex
ecological system has not been seriously studied, but the ecology
is believed to be in a dynamic state, not at a stationary state.
This diversity and dynamics are also seen in our population dynamics
showing homeochaos.  It is strongly hoped that the population dynamics of
rain forest species is measured soon. When this is done, we believe the
dynamics will be found to be homeochaotic. We also
believe that the mutation rate itself will be found to be
larger than the normal level.  If this is the case,
we may assume that the coupling among species is effectively larger
than in temperate zones.
We may also hope that our homeochaos is
important in a complex network system in general;
for a dynamical network system with many units evolving
according to some inherent dynamics.
Such examples may include neural systems, computer networks,
economics, and sociology.
Our homeochaos provides a key principle for the formation of
cooperation and the dynamic stability required in such systems.
\section{Source of Novelty and Growth of Diversity : Open Chaos}
In the previous section we have studied the maintenance of diversity.
How about its creation?  In \S 3,
we have discussed a possible theoretical basis for the origin of diversity.
The most typical origin of diversity is seen in cell differentiation.
The formation and maintenance of a society of differentiated
cells \cite{somatic} is also important for the origin of multicellular
organism.
By cell division, each cell reproduces itself with differentiation
and forms a network of cell society.
Is the dynamic clustering
mechanism in \S 3 relevant to cell differentiation?
To consider this problem, quite remarkable experimental results
are reported \cite{Yomo}:
E-coli with identical genes can split into several groups
with different enzymatic activities. Even prokaryotic cells with
identical genes can be differentiated there. Furthermore,
cells are under liquid culture, thus they are in an identical environment.
This experimental result must be surprising to molecular biologists and
also to those who study differentiation in the context of spatial
pattern formation ( e.g., along the line of Turing instability):
Neither genetic nor spatial information is essential to
the differentiation in the experiment.
On the other hand, the experimental result may not be so surprising
in the light of the dynamic clustering discussed in \S 3.
Nonlinear metabolic reaction is involved in each cell,
as well as nonlinear interaction with the soup.  Cells interact globally
with all other cells through the soup.  Thus
it is possible to expect that cell's chemical oscillations differentiate and
form some groups.
Recently Yomo and the author proposed a model of cell differentiation
based on the idea of dynamic clustering in \S 3 \cite{KKTY}.
The model consists of
metabolic reaction, active transport of chemicals
from a medium (soup), and cell division.  The metabolic reaction is
nonlinear
due to the feedback mechanism of catalytic reaction.
The interaction of cells is global, due to the competition for
taking chemicals (resources to produce enzymes and DNA) from the
medium.
Cell division is assumed to follow the chemical activities in each cell.
The division speed of a cell is assumed to be proportional to
its average chemicals included therein.
{}From a dynamical systems point of view, the model has a novel feature,
not included in the globally coupled map.  Here the number of degrees of
freedom varies via cell division.  When a new cell is born, we need
additional degrees of freedom to indicate the cell's state.
Numerical results of the model show that there are three successive stages
in the growth of the number of cells: coherent growth,
dynamic clustering, and fixed cell differentiation (see Fig.5).
At the first stage, oscillations of chemicals of all cells
are synchronized, and they divide synchronously.
Hence the number of cells increases as 1,2,4,8,16, $\cdots$).
The second stage starts when the oscillations of chemicals lose their
synchronicity, and the dynamic clustering of \S 3 is observed.
At the last stage, some (active) cells start to have more chemicals
than others. Disparity in chemical activities is observed.
The speed of division of active cells is at least $10^2$ times
faster than other cells.
The active cells may correspond to germ cells, while
others correspond to somatic cells.  Here, somatic cells are
also differentiated according to the concentration of contained chemicals.
The oscillations of chemicals at the third stage sustains weak chaos.
We note that the third stage is robust against external perturbations.
Since all cells compete for resources from the medium, the above
results may be interpreted as follows.
At the second stage, cells form a time sharing system for resources,
by the clustering of oscillations, while the differentiation
between poor and rich cells is formed spontaneously at the third stage.
The emergence of the third stage is rather new, unexpected
from the dynamic clustering in \S 3.
Indeed, we may introduce a new concept,  which we call ``open chaos."
We propose open chaos as a novel
and general scenario for systems with growing numbers of elements.
By the active transport dynamics of chemicals, the
difference between two cells can be amplified, since a
cell with more chemicals is assumed to get even more.
Tiny differences between cells can grow exponentially
if parameters satisfy a suitable condition.  A grown cell
is divided into two, with an (almost) equal partition of the contained
chemicals.  This process looks quite similar to chaos in the
Baker's transformation, involving stretching (exponential growth) and
folding (division).
One difference between our cell division mechanism
and chaos is that phase space itself changes after a division in
the former, while the orbit comes back to the original phase space
in the stretching-folding mechanism of chaos.
Our ``open chaos" concept is a novel and  general mechanism of instability
and
irregular dynamics in a system with growing phase space.  In
studies of artificial life, the term ``open ended evolution"
often refers to a dynamics whose phase space attains more dimensions
with the appearance of new species, strategies, and so forth.
Open chaos provides a mechanism for the way in which chaotic instability
in dynamical systems can trigger the expansion of the
dimension of phase space. It is interesting to
extend the present open chaos to areas studied in connection with open
ended evolution, such as economics, sociology, game theory, and so on.
\section{Beyond Top-down and Bottom-up Approaches}
There have been long debates between top-down and bottom-up
approaches in artificial intelligence and neural networks.
In the bottom-up approach some kind of ``order parameters"
constructed from a lower level gives a higher level,
related with some macroscopic behavior. In the top-down approach,
a few instructions send messages to lower-level elements.
Of course it is possible to include a weak feedback between top/bottom
levels, starting from each approach. An example is a simulation of ants
with pheromone.  The dynamics of lower-level units (ants' motion) leads to
a collective field of pheromone, which governs the motion of the
lower-level units.
Since the dynamics of lower-level units is governed by the higher-level
dynamics, this scheme is essentially
analogous with the Prigogine's dissipative structure \cite{Prigogine}
or Haken's slaving principle \cite{Haken}.
In these approaches, it is assumed that the top level is
represented by small number of degrees of freedom, while
the bottom level may involve a huge number of  degrees  of freedom.
Furthermore, relationships between elements
are fixed.  Although it is possible to include a nontrivial dynamics at a
macroscopic level, the behavior of each element is passive and
totally susceptible to a higher-level.
Our network of chaotic elements provides a different mechanism,
in the following sense: first, the top level is not necessarily
represented by only a few degrees of freedom; second, relationships
between
elements at the lower level can dynamically change; third,
elements are not passive, but are active and dynamical \cite{Chris2}.
The first point may seem just a complication at first glance, but this
is not necessarily so.  Often motion governed by few degrees of freedom
emerges, which however, does not last forever due to the second point
(dynamic change of relationships). Again high-dimensional motion
comes back, until another structure emerges.  This mechanism, described
as chaotic itinerancy in \S 3, is essential to replace the top-down
and bottom-up approaches.
In a network of chaotic elements, for example,
the order at the top level is destroyed by chaotic revolt against the
slaving principle \cite{GCM}, in contrast with passive elements
in traditional approaches.
In the population dynamics model by Ikegami and the author (mentioned
in
\S 5), the higher-level corresponds to the survival of species as a
collection of species.  The higher level emerges from the bottom level, but
it is not necessarily represented by few degrees of freedom.
\section{Conclusion}
Summing up, we have discussed a chaotic scenario of evolution,
which allows for the dynamical change of relationships of units,
and spontaneous formation and destruction of upper levels.
The origin and maintenance of complexity and diversity are
explained through this scenario.  We hope that chaos can remain a source
of
complexity, novelty, and diversity in the studies of artificial life.
\section{Acknowledgements}
I am grateful to
Ichiro Tsuda , Takashi Ikegami, Tetsuya Yomo, Yukito Iba,
Junji Suzuki, Kei Tokita, and Tomoyuki Yamamoto
for continual criticism and stimulating discussions.
I would also like to thank Chris Langton for
critical reading of the manuscript, and discussions on
his collectionism.
The work is partially supported by
Grant-in-Aids for Scientific
Research from the Ministry of Education, Science, and Culture
of Japan.

\pagebreak
Figure Caption
\vspace{.1in}
Fig.1:
Emergent landscape: Average score for the players with
parameters within $[a_i, a_i +\Delta]$ is plotted
with $a_i=-1+i\times \Delta$. $t_{trs}=255$.  The number of players is
fixed
at 200.
a) The mutation rate  $\mu =0.1$, and $T=32$. The bin size $\Delta
=0.001$.
Sampled for time steps from 1000 to 1500, over all players. ( See for
details \cite{KKJS}, from which the figure is adapted).
b) $\mu =0.001$, and $T=128$.  The bin size $\Delta =0.002$.
Sampled for time steps from 750 to 1000, over all players.
\vspace{.1in}
Fig.2:
Average score of the game vs. Lyapunov exponents.
Simulation is carried out with $\mu =0.05$, $t_{trs}=255$,
and $T=128$ by fixing the population of birds at 200.
Average scores are obtained from the
histogram of  Lyapunov exponents, for which we use a bin
size of 0.01 for $-1<\lambda<1$,
while it is set at 0.1 for $\lambda<-1$ ( since the sample there
is rather sparse).
Sampled over
time steps from 500 to 750 over all players ( whose number is
fixed at 200).
Fig.3: Space-time diagram for the coupled map lattice on a
hypercubic lattice with $k=5$ (i.e., $N=2^5$).
For local dynamics $f(x)=1-1.52x^2$ is adopted, while
the coupling strength $\epsilon$ is set at 0.3.
On the corresponding pixel at a given time and element,
a bar with a length
proportional to $(x_n(i,j)-0.1)$ is painted if $x_n(i,j)>.1$.
Every  4th time step is plotted from 10000 to 12000.
Elements are aligned according to its binary representation;
i.e., $0=00000,1=00001$,$2=00010$, $\cdots$, $63=11111$.
At the stage A, elements split into two clusters $0*****$ and $1****$,
while they split into $*0***$ and $*1***$ at the stage B,
$**0**$ and $**1**$ at the stage C,  $0*****$ and $1****$ at the stage D,
$****0$ and $****1$ at the stage E, and again into
$0*****$ and $1****$ at the stage F.
\vspace{.1in}
Fig.4: Oscillation of total populations for hosts (solid line) and parasites
(dashed line), in the model in \cite{homeo}. Initial chaotic
oscillations with large amplitudes are suppressed
simultaneously with the increase of averaged mutation rate
around time $=$ 2300.  Up to the time, the population dynamics is
a direct product of low-dimensional chaos, and includes a violent change
of populations.  After the time, the oscillation amplitude is much weaker
( although chaotic), where homeochaos is attained with the increase of
the mutation rates.  See for details \cite{homeo}.
\vspace{.1in}
Fig.5:  Overlaid time series of a chemical at each cell.
As the number of cells grows, the oscillation starts
and clustering emerges at the stage II.  At the stage III,
disparity of the chemical is clearly seen.
See for details \cite{KKTY}, from which the figure is adapted.

\begin{thebibliography}{999}
\bibitem{equiv}
It is equivalent to
\begin{math}
y_{n+1}(i)=f[(1-\epsilon )y_{n}(i)+\frac{\epsilon }{N}\sum_{j=1}^N y_{n}(j)],
\end{math}
with the aids of transformation $y_n (i)=f(x_n (i))$.
In this form, one can see clear correspondence with
neural nets: If one chooses a sigmoid function (e.g.,$tanh(\beta x)$)
as $f(x)$ and a random or coded coupling $\epsilon _{i,j}$,
a typical neural nets is obtained.
\bibitem{HCM-com}
Here we use an identical dynamics for all species.  To include the fitness,
one can adopt element-dependent dynamics $f_i (x)$, instead.  In this case
we can see some effects of clusterings, too.
\end{thebibliography}
\end{document}